%% file: main.tex
\documentclass[]{spie}  

\usepackage[square,numbers,sort&compress,comma]{natbib}
\usepackage{amsmath,amsfonts,amssymb}
\usepackage{graphicx}
\usepackage[export]{adjustbox}
\usepackage{caption}
\graphicspath{{./}{./figures/}}
\usepackage[colorlinks=true, allcolors=blue]{hyperref}
\usepackage{threeparttable}
\usepackage[usenames,dvipsnames,svgnames]{xcolor}

\newcommand{\CHECK}[1]{{#1}}

\newcommand{\eq}{\begin{equation*}}
\newcommand{\qe}{\end{equation*}}

\usepackage{aas_macros}

\input{commands}

\title{Characterization of Skipper CCDs for Cosmological Applications}

\reportnum{-2.5}{FERMILAB-CONF-20-579-AE-LDRD}

\author[a,b,*]{Alex Drlica-Wagner}
\author[b]{Edgar Marrufo Villalpando}
\author[a]{Judah O'Neil}
\author[a]{Juan Estrada}
\author[c]{Stephen Holland}
\author[a,b]{Noah Kurinsky}
\author[d,e]{Ting Li}
\author[a]{Guillermo Fernandez Moroni}
\author[a]{Javier Tiffenberg}
\author[a,f]{Sho Uemura}

\affil[a]{\small Fermi National Accelerator Laboratory, Batavia, IL 60510, USA}
\affil[b]{\small Kavli Institute of Cosmological Physics, University of Chicago, Chicago, IL 60637, USA}
\affil[c]{\small Lawrence Berkeley National Laboratory, One Cyclotron Rd, Berkeley, CA 94720, USA}
\affil[d]{\small Observatories of the Carnegie Institution for Science, Pasadena, CA 91101, USA}
\affil[e]{\small Department of Astrophysical Sciences, Princeton University, Princeton, NJ 08544, USA}
\affil[f]{\small Raymond and Beverly Sackler School of Physics and Astronomy, Tel-Aviv University, Tel-Aviv 69978, Israel}

\authorinfo{$^*$~kadrlica@fnal.gov}

\begin{document} 
\maketitle

\begin{abstract}
We characterize the response of a novel $250\um$ thick, fully-depleted Skipper Charged-Coupled Device (CCD) to visible/near-infrared light with a focus on potential applications for astronomical observations.
We achieve stable, single-electron resolution with readout noise $\sigma \sim 0.18 \ermspix$ from 400 non-destructive measurements of the charge in each pixel.
We verify that the gain derived from photon transfer curve measurements agrees with the gain calculated from the quantized charge of individual electrons to within $< 1\%$.
We also perform relative quantum efficiency measurements and demonstrate high relative quantum efficiency at optical/near-infrared wavelengths, as is expected for a thick, fully depleted detector.
Finally, we demonstrate the ability to perform multiple non-destructive measurements and achieve sub-electron readout noise over configurable sub-regions of the detector. 
This work is the first step toward demonstrating the utility of Skipper CCDs for future astronomical and cosmological applications.
\end{abstract}

\keywords{Skipper CCD, sub-electron noise, photon counting detector, spectroscopy}

\section{INTRODUCTION}
\label{sec:intro}  

Charge-coupled devices (CCDs) have led to transformative advances in photon detection for many industrial and scientific applications \cite{Boyle:1970,Amelio:1970,damerell:1981,Janesick:2001}.
CCD sensors rely on the photoelectric effect to absorb incident photons in a silicon substrate and generate electron-hole pairs~\cite{Amelio:1970}.  
Optical and near-infrared photons produce one electron-hole pair, while more energetic photons can produce multiple electron-hole pairs. 
CCDs have been widely used in astronomy for decades, and current detectors can have quantum efficiencies $> 90\%$, dynamic ranges of $\roughly 10^5$, and pixel sizes of $\roughly 10 \um$ \cite{Janesick:2001}.

Precision measurements with CCDs have been limited by the electronic readout noise that is added to the CCD video signal by the output amplifier \cite{Janesick:2001}.
While correlated double sampling dramatically reduces high-frequency readout noise \cite{CDS:1050448}, low-frequency readout noise has remained a fundamental limitation for precision single-photon/electron counting in CCDs. 
In conventional scientific CCDs, low-frequency readout noise results in root-mean-squared (rms) variations in the measured charge per pixel of $\sim 2 \ermspix$ \citep{Holland:2003,Janesick:2016,Haro:2016xoz,Bebek:2017,Aguilar-Arevalo:2019}. 
Janesick et al.\ \cite{Janesick:1990,Janesick:1992} proposed that low-frequency readout noise could be reduced by using a floating gate output stage \citep{Wen:1974} to perform repeated measurements of the charge in each pixel.
This repeated measurement technique was implemented in the early 1990's in a device called a ``Skipper CCD'' \cite{Janesick:1990,Chandler:1990}; however, the noise performance of these detectors deviated from expectations below $\roughly 0.5 \ermspix$ \citep{Janesick:1990}.
Recently, Tiffenberg et al.\ \cite{Tiffenberg:2017} demonstrated that Skipper CCDs designed at Lawrence Berkeley National Laboratory and fabricated by Teledyne DALSA Semiconductor could implement a floating gate output stage, a small-capacitance sense node, and isolation from parasitic noise sources to perform thousands of independent, non-destructive measurements of the charge in a pixel.
This architecture allows for the drastic reduction of low-frequency readout noise, and recent devices have achieved a readout noise of $< 0.04 \ermspix$ \cite{Cancelo:2020}.
This new generation of Skipper CCDs have found broad applications as dark matter detectors \cite{Crisler:2018,Abramoff:2019,Barak:2020} and for characterizing the material properties of silicon \cite{Rodrigues:2020}.

Astronomical observations can also benefit from reduced readout noise.
In particular, observations of faint objects in the low-signal, low-background regime are currently limited by detector readout noise. 
One of the major emphases of the 2010 Decadal Review in Astronomy was direct imaging and spectroscopy of terrestrial exoplanets in the habitable zones of nearby stars \cite{Astro2010}. 
These experiments have a stated requirement on detector readout noise at the level of $0.1 \ermspix$ in the visible/near-infrared wavelengths \cite{Crill:2017,ExoPlanet:2019}.
Another application of low-noise detectors is to multi-object spectroscopy of faint stars and galaxies \citep{DESI:2016a}.
In particular, medium to high resolution spectroscopy at blue wavelengths has low sky-background levels and significant gains could be achieved by reducing the readout noise to $\roughly 0.5 \ermspix$.
Finally, high-cadence searches for short duration transients (e.g., fast radio bursts) could benefit from reduced detector readout noise \cite{Richmond:2020,Tingay:2020}.
Skipper CCDs are an attractive approach to reducing readout noise since they are expected to possess many of the beneficial characteristics of conventional astronomical CCDs (i.e., stability, linear response, large dynamic range, high quantum efficiency, and radiation response).
In this paper, we demonstrate the performance of a Skipper CCD at visible/near-infrared wavelengths in anticipation of using Skipper CCDs for astronomical observations.

\section{MOTIVATION}

To quantify the benefit of reduced readout noise on astronomical observations, we review the formalism for calculating the expected signal-to-noise ratio as a function of readout noise.
In an astronomical observation, the total noise per pixel per exposure can be broadly decomposed into two components,
\begin{equation}
\sigma^2 = \sigma_{\rm shot}^2 + \sigma_{\rm read}^2.
\end{equation}
Here, $\sigma_{\rm shot}$ is the Poisson shot noise contributed by the source of interest, contaminating background light (e.g., scattered light, sky background, zodiacal light, etc.), and detector dark current.
The total shot noise can be derived from the rate, $r$ (\epixsec), for each contributing component and the integration time, $t_{\rm exp}$ (s), of the exposure,
\begin{align}
\sigma_{\rm shot}^2 &= (r_{\rm src} + r_{\rm bkg} + r_{\rm dark}) t_{\rm exp}.
\end{align}
In contrast to the shot noise, the readout noise is independent of the exposure time and is generally expressed in terms of the rms fluctuations in the number of electrons per pixel (\ermspix).

Observationally, the counts from a source will be spread over several pixels. 
Assuming uncorrelated pixel measurements, the noise components add in quadrature.
Thus, the total noise integrated over the pixels of interest will be,
\begin{align}
\Sigma_{\rm tot}^2 = N \sigma_{\rm tot}^2 &= N (r_{\rm src} + r_{\rm bkg} + r_{\rm dark}) \texp + N \sigma_{\rm read}^2 \\
 & = (R_{\rm src} + R_{\rm bkg} + R_{\rm dark}) \texp + N \sigma_{\rm read}^2,
\end{align}
where in the second line we have defined the total source, background, and dark contributions integrated over pixel elements ($R_{\rm src}, R_{\rm bkg}, R_{\rm dark}$), since these numbers are commonly provided by astronomers.

The sensitivity to a given source is generally expressed by the signal-to-noise ratio (\SNR), which is defined as the ratio between the number of counts contributed by the source and the total noise in the observation,
\begin{align}
\SNR = \frac{R_{\rm src} \texp}{\Sigma_{\rm tot}}
     = \frac{R_{\rm src} \texp}{\sqrt{ (R_{\rm src}+R_{\rm bkg} + R_{\rm dark})\texp + N \sigma_{\rm read}^2}}.
\label{eqn:snr}
\end{align}
From this formulation, it is clear that the total noise will be dominated by the readout noise only in cases where $N \sigma_{\rm read}^2 \geq (R_{\rm src} + R_{\rm bkg} + R_{\rm dark}) \texp$.
When the readout noise dominates then $\SNR \propto \texp$ as opposed to the familiar behavior $\SNR \propto \sqrt{t_{\rm exp}}$.
Most astronomical observations are designed to avoid the readout-noise dominated regime.
In particular, for bright sources $R_{\rm src} \gg R_{\rm bkg} \gg R_{\rm dark}$, and any moderate exposure time yields $R_{\rm src} \texp \gg N\sigma_{\rm read}^2$.
In contrast, for faint sources $R_{\rm src} < R_{\rm bkg}$ and achieving a reasonable signal-to-noise requires a long exposure time such that $R_{\rm bkg} \texp > N \sigma_{\rm read}^2$.
Thus, the readout-noise dominated regime occurs only in scenarios where both the signal and background rates are small and the exposure time is constrained such that longer integration times are not possible. 
Such observations are necessarily in the low signal-to-noise regime.
For ground-based imaging of the static sky, Poisson shot noise from the source of interest and the sky background generally dominate over the readout noise of current detectors. 
However, for spectroscopic measurements from the ground (especially at short wavelengths where the sky background is greatly reduced), space-based imaging and spectroscopy of faint sources (e.g., extra-solar planets), and very high-cadence imaging for short duration transients (e.g., fast radio bursts), readout noise can be an important contribution to the overall noise budget.
In these cases, reducing readout noise can achieve higher signal-to-noise with shorter observation times.

\begin{figure}[t!]
\includegraphics[width=0.49\textwidth]{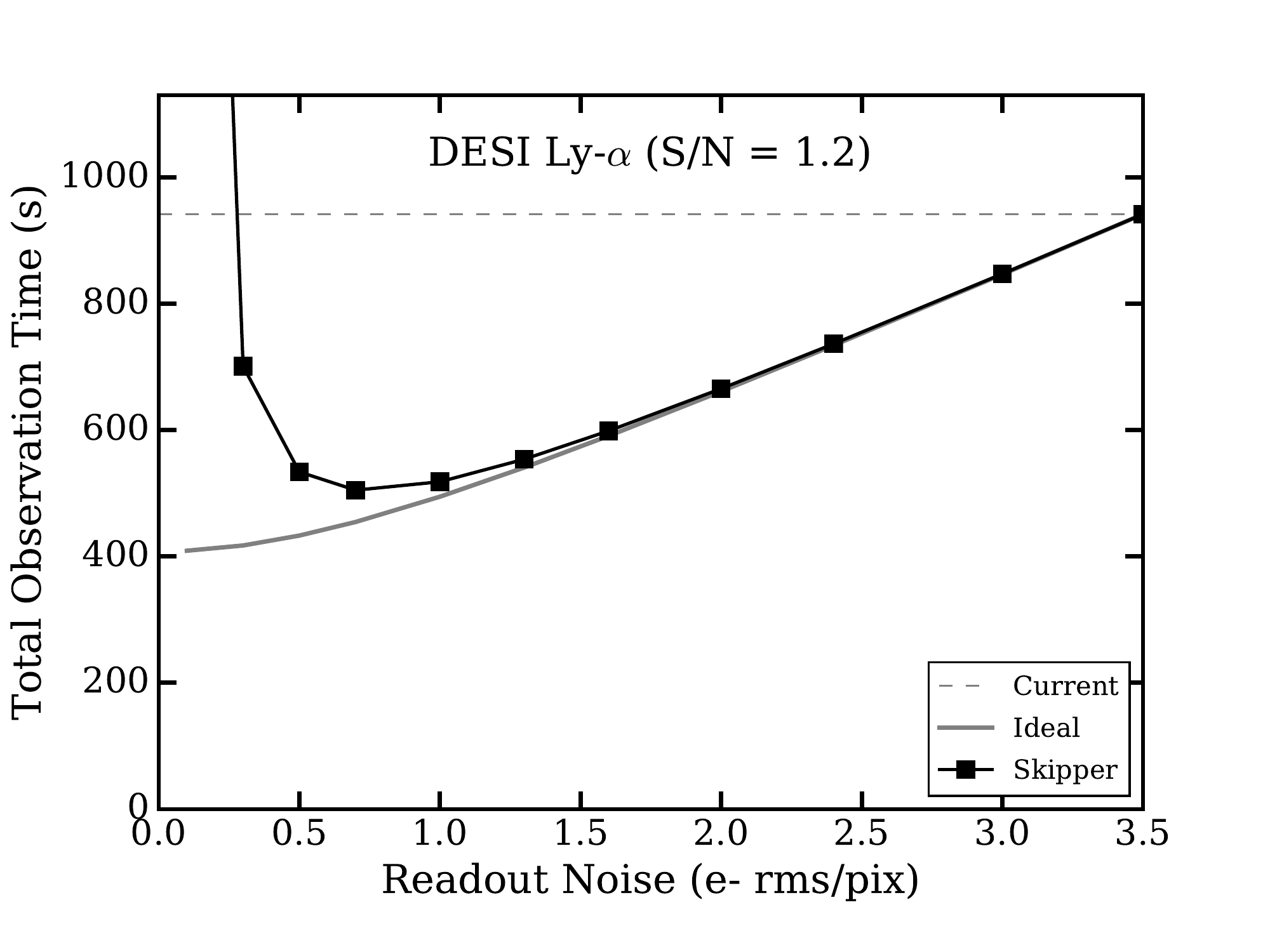}
\includegraphics[width=0.49\textwidth]{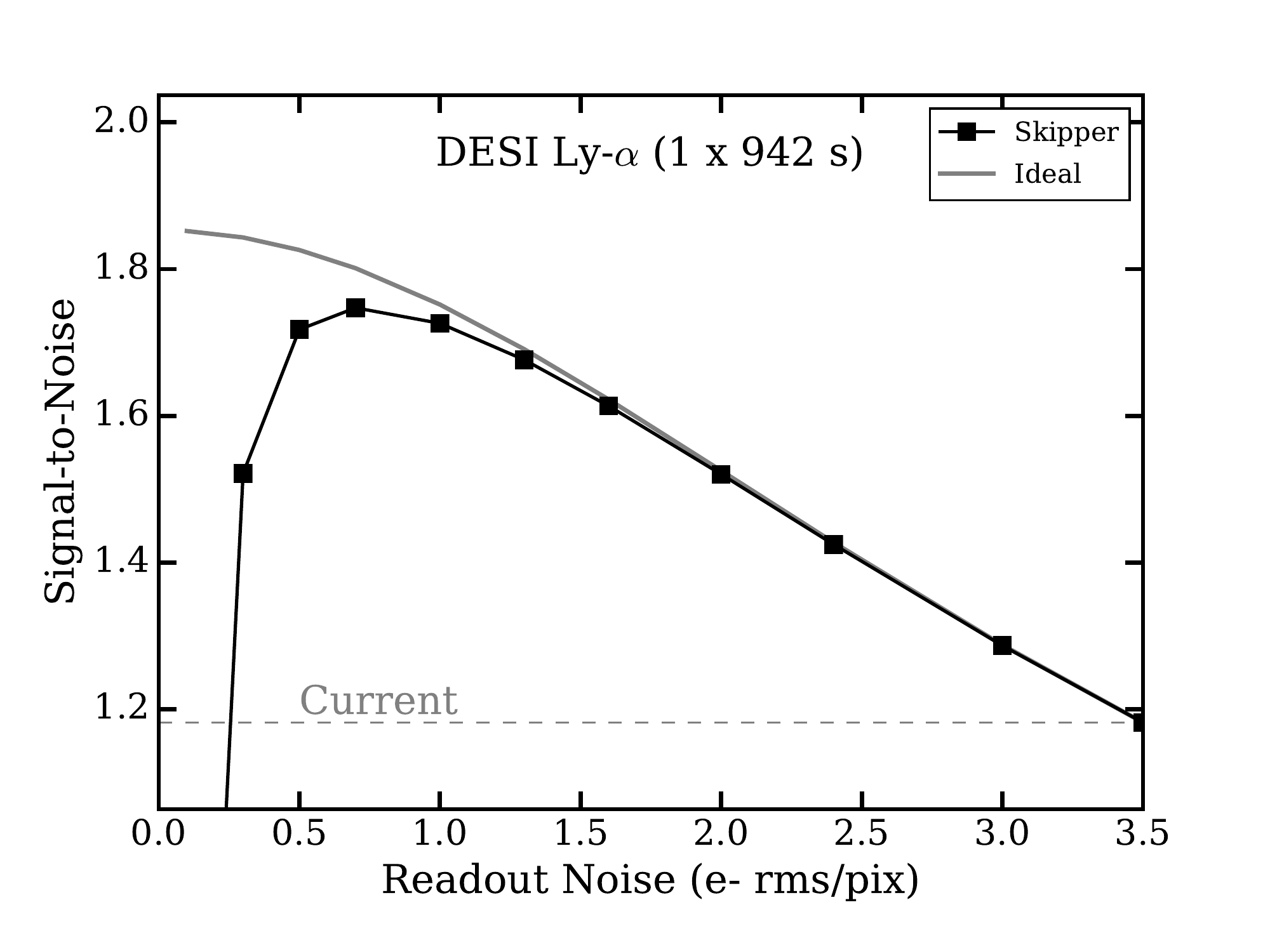}
\caption{Expected improvement in DESI \Lya observations from reduced readout noise. The gray line shows the ideal case where readout noise is reduced without increasing readout time, while the black line shows the expected performance of a Skipper CCDs that reduces readout noise over 5\% of the detector area. Left: Decrease in the required observation time (the sum of the exposure time and readout time) to reach a fixed \SNR. Right: Increased \SNR at fixed observation time as a function of readout noise.} 
\label{fig:snr}
\end{figure}

As an example, we explore the signal-to-noise improvement possible for ground-based observations of the \Lya forest using a multi-object spectrograph. 
We compare against the state-of-the-art system assembled for the Dark Energy Spectroscopic Instrument (DESI) \cite{DESI:2016a,DESI:2016b}.
Readout noise is an important contributor to the noise budget of DESI \Lya observations and dominates at wavelengths $< 3700 \angstrom$. 
We estimate the DESI \Lya performance for a zenith observation during dark time with 0.92 arcsec seeing and an exposure time of $900\second$.
We use realistic simulations to estimate the \Lya signal and sky background contributions (Julien Guy, private communication).
We find that the number of electrons contributed by a quasar with $g = 22$ (AB) per $1\angstrom$ flux bin is $15 \e/\angstrom$, while the dark sky background contributes $40 \e/\angstrom$ (accounting for fiber size).
The effective number of CCD pixels contributing to the noise of a flux bin of width $1 \angstrom$ is derived from extracted spectra using ZEMAX point-spread model and is found to be $N = 7.8 \pix/\angstrom$.
Converting these numbers into the quantities used in \eqnref{snr}, we find $\texp = 900\second$, $r_{\rm src} = 2.1\times 10^{-3} \epixsec$, $r_{\rm bkg} = 5.7 \times 10^{-3} \epixsec$, $\sigma_{\rm read} = 3.5 \ermspix$, and $N = 7.8 \pix$.
Evaluating \eqnref{snr} yields a \SNR per exposure of $\SNR = 1.2$; extending to the four exposures expected from DESI yields $\SNR = 2.4$.

Skipper CCDs can reduce the readout noise to an arbitrarily low level; however, this improvement comes at the cost of increased readout time to perform multiple measurements of the charge in a pixel. 
Since time used on readout could alternatively be used to collect additional photons, there is a natural maximum improvement that can be achieved by reducing noise levels with Skipper CCDs.
We quantify this improvement in performance through two metrics: (1) the reduction in observation time to reach fixed \SNR, and (2) the improvement in \SNR at fixed observation time.
For both cases, we define the observation time as the sum of the exposure time and the readout time. 
Skipper CCD readout time per pixel scales nearly linearly with the number of samples, and assuming that only some fraction, $f$, of the detector pixels would need to be read out with improved \SNR, we calculate the observation time as,
\begin{equation}
\tobs = \texp + ( 1 - f) \tread + f N_{\rm samp} \tread.
\end{equation}
Assuming a targeted detector region of $f = 5\%$, we can calculate the optimal readout noise as to maximize the \SNR at fixed observation time or alternatively, minimize the observation time at fixed \SNR (\figref{snr}).
In both cases we find that the optimal readout noise is $\sigma_{\rm read} \sim 0.5 \ermspix$, which provides a factor of 48\% increase in the \SNR at fixed observation time, or alternatively achieves the DESI \Lya $\SNR = 1.2$ per exposure in 54\% of the observation time.
We use this gain in performance to motivate further study of Skipper CCDs for astronomical applications.

\section{EXPERIMENTAL SETUP}

\begin{figure} [t!]
   \begin{center}
   \begin{tabular}{c} 
   \includegraphics[width=0.44\textwidth]{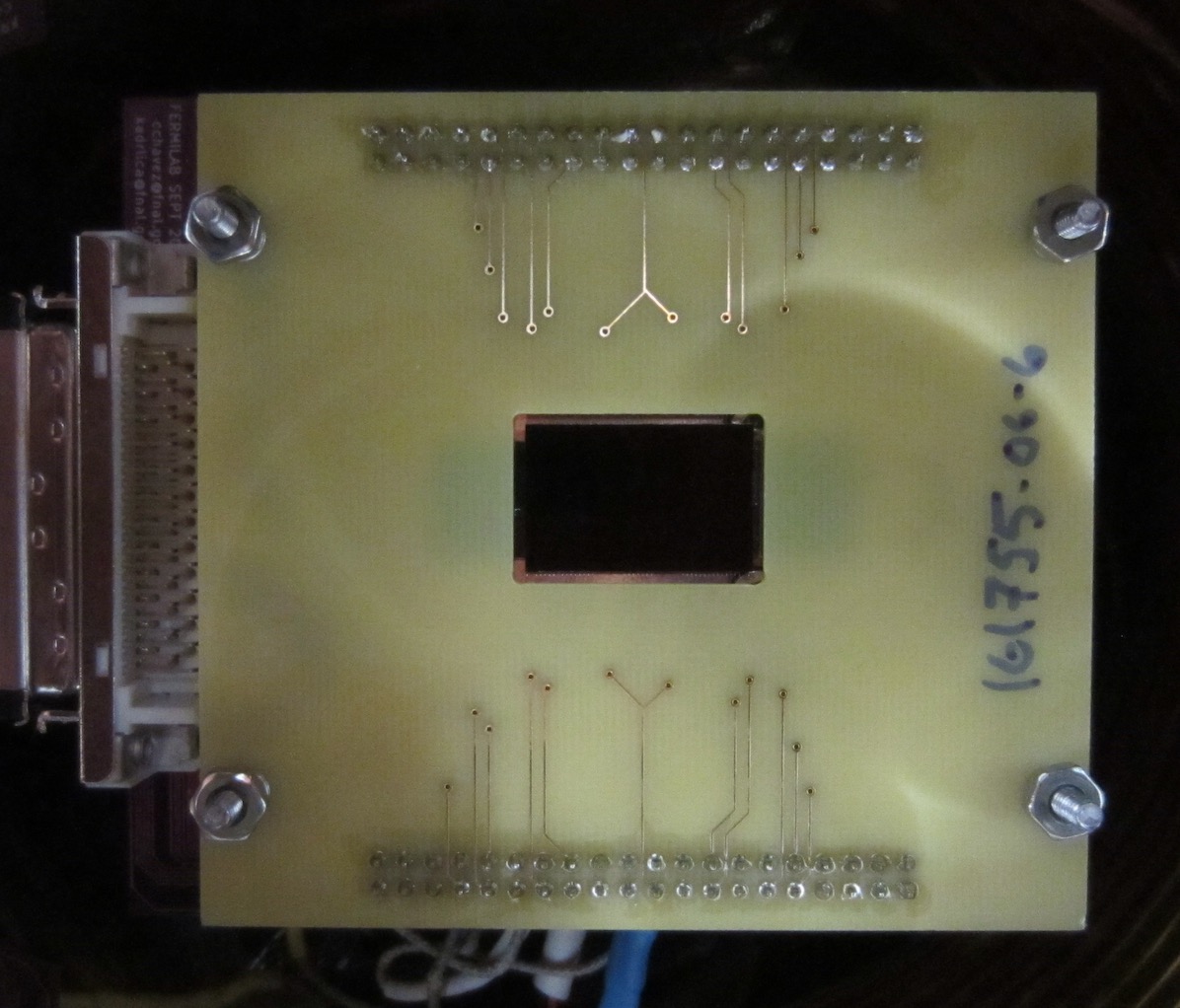}
   \includegraphics[width=0.55\textwidth]{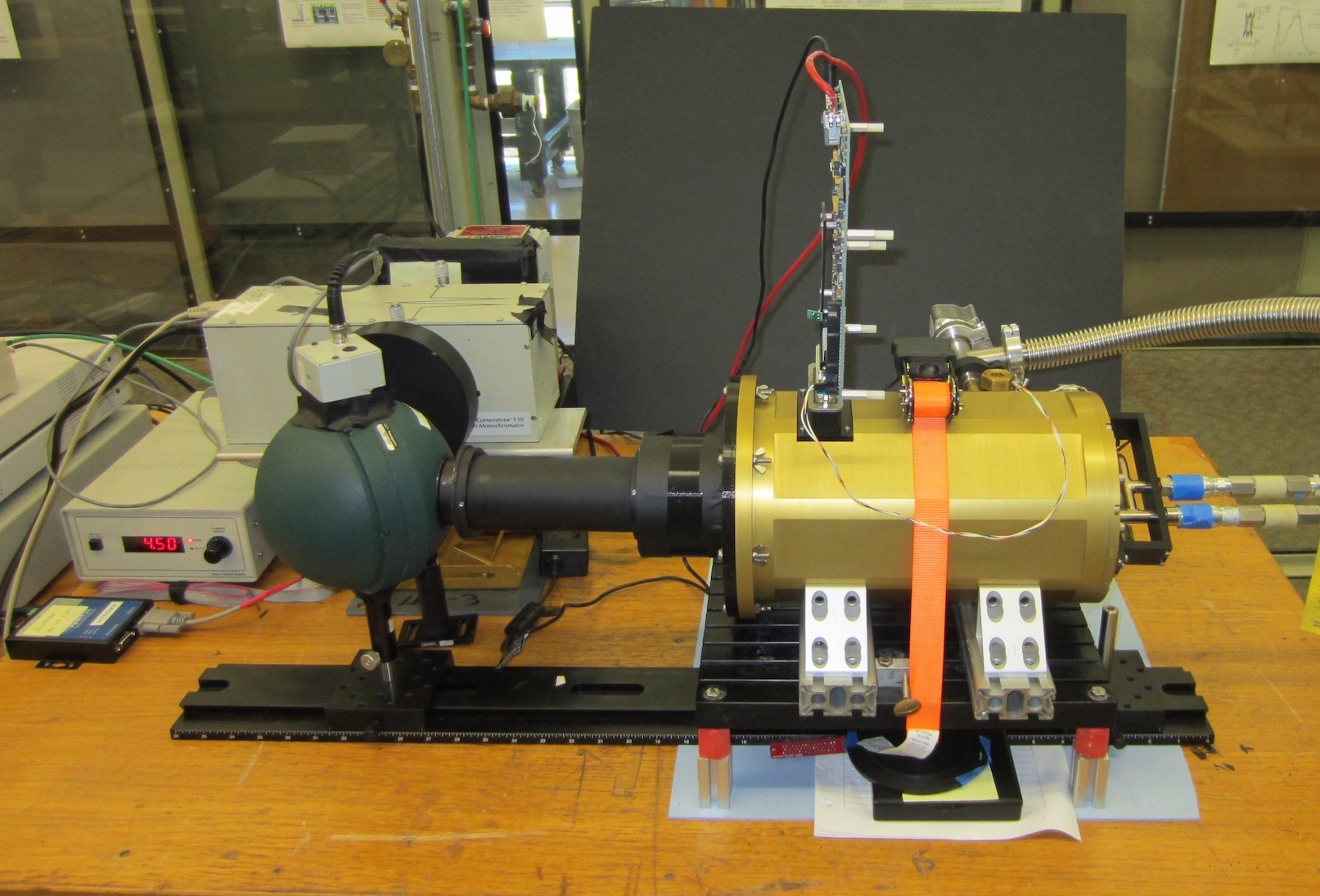}   
   \end{tabular}
   \end{center}
   \caption[example] 
   {\label{fig:setup} Left: Backside-illuminated Skipper CCD mounted in a picture frame adapter.
     Right: Experimental setup including a closed-cycle optical dewar (gold; right) housing the Skipper CCD with attached LTA readout board and integrating sphere with attached photodiode (green; left). Behind the integrating sphere are a mechanical shutter, monochromator, filter wheel, and tungsten halogen lamp. 
   }
\end{figure} 

\subsection{Detectors and Packaging}
\label{sec:detectors}

The detector studied here is a $p$-channel Skipper CCD fabricated on high resistivity ($\sim$5~k$\Omega\cm$), $n$-type silicon (\figref{setup}). 
The CCD is $250\um$ thick and composed of $15 \um \times 15 \um$ square pixels arranged in a $1248 \times 724$ array (\tabref{detector}).
This detector was produced as part of an R\&D run exploring different floating gate configurations, and thus has several floating gate layout configurations: U1 ($15\um \times 4\um$), L1 ($24\um \times 9\um$), U2 ($24\um \times 4\um$), L2 ($24\um \times 4\um$).
Our studies focus on the performance of the \CHECK{U2} configuration, which was found to be the best performing on this detector.
This detector was designed at Lawrence Berkeley National Laboratory and fabricated by Teledyne DALSA Semiconductor. 

The detector was packaged for backside illumination using on a ``picture frame'' adapter board \cite{Derylo:2006}.
Because the front side of these R\&D CCDs were not passivated, special care was needed to avoid damaging the traces on the front side of the CCD during packaging.
The front side of the CCD was epoxied to a narrow, blank silicon substrate, which was itself epoxied to the printed circuit board (PCB) of the picture frame.
Double-sided tape was used to hold the CCD in place on the silicon substrate while the epoxy dried.\footnote{The use of tape during packaging results in non-uniformity of the pixel response that is visible at high light levels  \cite{Derylo:2006,Plazas:2014}. However, this packaging method is sufficient for our testing purposes.}
Wire bonding was performed between the pads of the CCD and the traces on the picture frame PCB.
The picture frame adapter was connected through another PCB to a 50-pin flexible cable.
This flexible cable is similar to the second-stage cable from DECam, containing a JFET source follower and an operational amplifier \cite{Flaugher:2015,Tiffenberg:2017}. 
The flexible cable is connected to a DB-50 vacuum feed-through.
CCD packaging and testing was carried out in the Silicon Detector Facility (SiDet) at Fermilab.

The sensor was fully depleted at a substrate bias voltage of $40$\,V.
An operating temperature of 140\,K was used to reduce the number of electrons promoted to the conduction band by thermal fluctuations (``dark current'')~\citep{Holland:1989,Holland:2003}.
Similar thick high-resistivity Skipper CCDs that are shielded from infrared radiation have a measured dark current of $\roughly 4 \times 10^{-3} \epixday$ at 150\,K and are extrapolated to have a dark current of $\roughly 2 \times 10^{-4} \epixday$ at 140\,K \cite{Barak:2020}.

\begin{table}[t]
\centering
\caption{\label{tab:detector}
Skipper CCD Detector \& Operating Characteristics
}
\begin{tabular}{l c c c}
\hline
Characteristic  & Value  & Unit\\
\hline \hline
Format & $1248 \times 724$ & pixels\\
Pixel Scale & 15 & $\um$ \\
Thickness & $250$ & $\um$\\
Number of Amplifiers & 4 & \\
\hline
Operating Temperature & 140 & K \\
Substrate Voltage & 40 & V \\
\hline
\end{tabular}
\end{table}

\subsection{Readout Electronics}
\label{sec:electronics}

Our detector is controlled by a Low Threshold Acquisition (LTA) board for Skipper CCDs \cite{Cancelo:2020}.
The LTA features an on-board FPGA that controls the programmable bias voltages, clock signals, video acquisition, telemetry, and the shipment of data and metadata from the board to a computer.
The FPGA controls the readout sequence to ensure precise synchronization of clock signals and ADC samples processing during the CCD readout.
Data and commands are transmitted entirely using UDP over IP through a gigabit Ethernet port. 
Voltages, integration windows, and clock sequences can be configured at run time, and software converts the output voltage stream to FITS images for subsequent processing.
The LTA hosts four video channels allowing the four CCD output channels to be read simultaneously.
More details on the LTA controller can be found in Cancelo et al. \cite{Cancelo:2020}.

\subsection{Mechanical System}
\label{sec:mechanical}
The Skipper CCD is installed in a closed-cycle vacuum dewar with a fused silica window manufactured by Infrared Laboratories Inc.
A copper block was machined and attached directly to the cold finger that penetrates the dewar.
The CCD in the picture frame adapter is mounted to the copper block, such that the blank silicon packaging substrate is kept in thermal contact with the copper block.
The system is cooled by a closed-system cryocooler.
A 40\,W cartridge heater is installed in the copper block and is controlled by a Lakeshore temperature controller to maintain a stable operating temperature of 140\,K.
We restrict the cool-down to a rate of 1K/min to prevent damage due to rapid thermal contraction of the CCD, silicon blank, or PCB.

The vacuum and cooling lines are electrically isolated from the CCD system signal ground to avoid injecting low-frequency noise into the CCD readout system. 
The LTA is powered by a +12V DC power brick, and the LTA connects to the readout computer through a standard Ethernet cable. 
The LTA data acquisition software is written in C++ and runs on a standard Linux operating system.
The LTA software can be used to set readout parameters (e.g., voltages, timing, clocking sequence) and get status information through the LTA on-board telemetry. 
The data acquisition software generates images in FITS file format for further analysis. 

\subsection{Optical System}
\label{sec:optical}
The CCD was illuminated through a standard set of optical equipment including a tungsten arc lamp, motorized filter wheel, monochromator, shutter, and integrating sphere (\figref{setup}).
This setup provided uniform illumination to the CCD surface with configurable wavelength and exposure time.
The shutter, filter wheel, and monochromator were controlled using GPIB through a TCL interface. 
Light intensity was measured independently by a NIST-calibrated photodiode connected to an auxiliary port on the illuminating sphere.
This allows us to measure the relative illumination; however, we do not have the ability to perform absolute photon flux measurements at the surface of the CCD.


\section{DETECTOR CHARACTERIZATION}

\begin{table}[t]
\centering
\begin{threeparttable}
\caption{\label{tab:performance}
Skipper CCD Detector Performance
}
\begin{tabular}{l c c c}
\hline
Characteristic  & Value  & Unit\\
\hline \hline
Readout Noise (1 sample) & {$3.57$} & $\ermspix$ \\
Readout Noise (200 samples) & {$0.25$} & $\ermspix$ \\
Readout Noise (400 samples) & {$0.18$} & $\ermspix$ \\
Readout Time (400 samples; 1 amp) & 50 & $\us$/pix/samp \\
Readout Time (400 samples; 4 amps) & 12.5 & $\us$/pix/samp \\
Saturation Level$^a$ & \CHECK{$\roughly34{,}000$} & \e \\
Gain & 162 & ADU/\e \\
Charge Transfer Inefficiency & {$1.9 \times 10^{-6}$} & \\
\hline
\end{tabular}
\begin{tablenotes}
      \small
      \item $^a$ Full-well capacity is likely much higher. See \secref{ptc} for details.
\end{tablenotes}
\end{threeparttable}
\end{table}

\subsection{Readout Noise}
\label{sec:readnoise}
The Skipper CCD provides the capability to dynamically adjust detector readout noise through multiple non-destructive measurements of the charge in each pixel (commonly referred to as the number of ``samples'' per pixel). 
For uncorrelated Gaussian readout noise, the measured charge in each pixel is given by the average of the individual measurements of that pixel. 
The readout noise on the average of the individual measurements is expressed as
\begin{equation}
    \sigma_N = \frac{\sigma_{1}}{\sqrt{N_{\rm samp}}},
    \label{eqn:noise}
\end{equation}
where $\sigma_{1}$ is the single-sample readout noise (the standard deviation of pixel values with a single charge measurement per pixel), and $N_{\rm samp}$ is the number of measurements per pixel \citep{Tiffenberg:2017}. 
We characterize the readout noise of our detector as a function of the number of samples, showing that it obeys the theoretical expectation in \eqnref{noise}.
We demonstrate sub-electron resolution and perform a direct measurement of detector gain (ADU/\e) from the values of pixels containing $0\e$, $1\e$, $2\e$, etc.
Results are collected in \tabref{performance}.

\subsubsection{Data Acquisition}

To quantify detector readout noise as a function of the number of samples, we performed multi-sample readout of the detector during dark and low-illumination exposures.
For each exposure, we read 450 columns from the first 50 rows of the detector with 400 samples per pixel. 
Each row contained 8 prescan pixels, 362 physical pixels, and 80 virtual overscan pixels.
The overscan pixels were used to measure the detector readout noise and correct the bias level of the physical pixels.
Between each multi-sample exposure, we took a single-sample exposure to ``clean'' additional charge from the CCD (i.e., charge stored in rows that were not read).


\subsubsection{Results}

\begin{figure}[t!]
    \centering
    \includegraphics[width=0.49\textwidth]{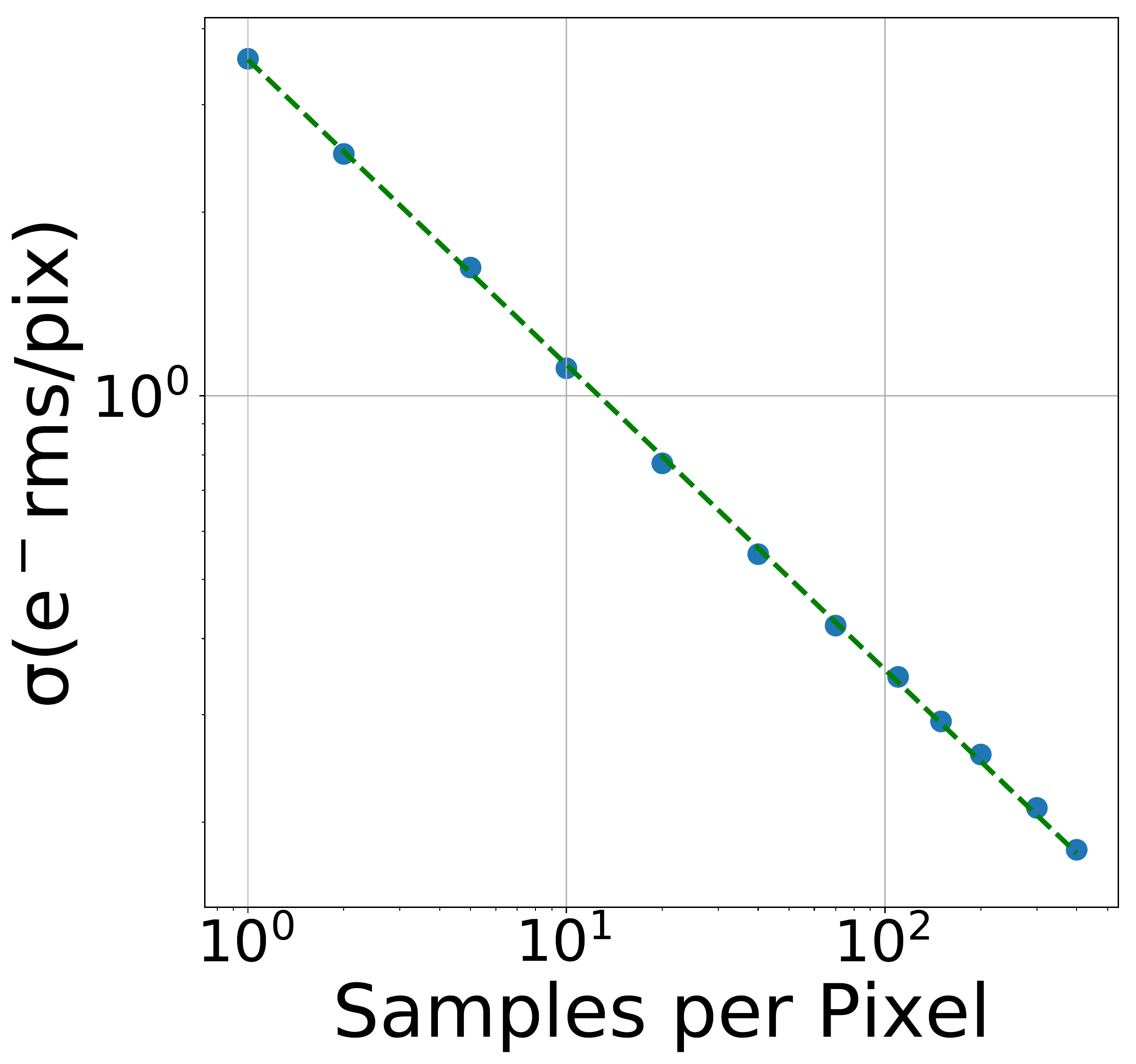}
    \includegraphics[width=0.49\textwidth]{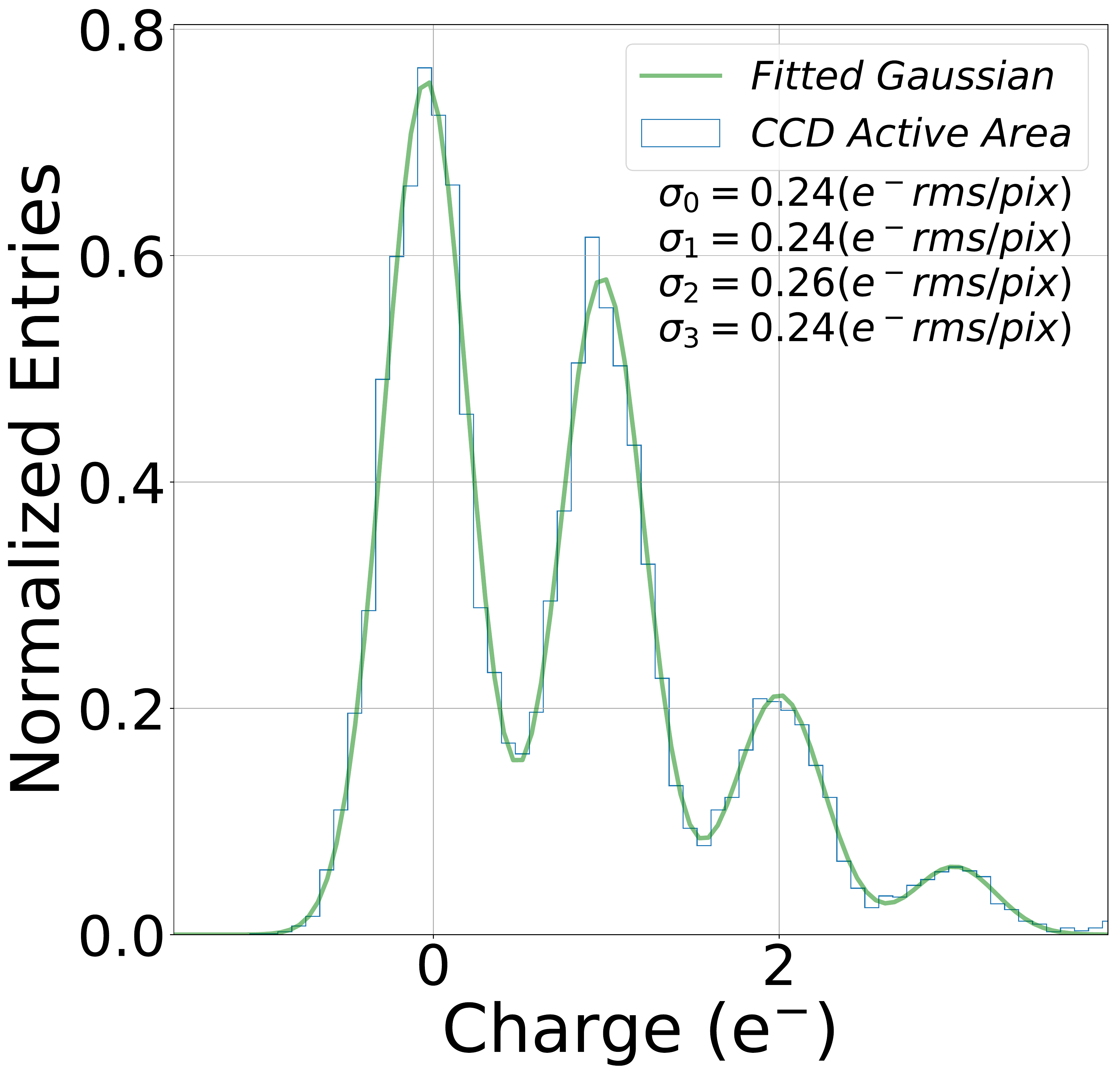}
    \caption{Readout noise performance of Skipper CCD. Left: Readout noise as a function of the number of samples for the Skipper CCD (blue points). Performance closely follows the $1/\sqrt{N_{\rm samp}}$ predicted from the central limit theorem (green dashed line). Right: Histogram of pixel values calculated from the average of 200 measurements per pixel (blue histogram).
Single-electron resolution is clearly achieved with a readout noise of $\sigma_{200} \sim 0.25 \ermspix$. 
The peaks in the distribution are well-fit by a multi-Gaussian model (green line).
The gain of the CCD can be directly measured from the spacing between peaks.}
    \label{fig:skipper}
\end{figure}

{\it Readout Noise:} We measure the readout noise performance of the detector from the distribution of overscan pixel values on a 400-sample image.
Due to the long readout time ($\roughly 450\second$ per image), a small light leak can lead to charge collection in the overscan pixels when they are shifted into the serial register. 
The presence of this excess charge leads to a broadening of the pixel value distribution, which is measurable at the very low readout noise achieved by the Skipper CCD.
To account for this charge, we model the pixel value distribution with a multi-Gaussian model and estimate the readout noise from the standard deviation of the Gaussian corresponding to the $0\e$ peak (see \figref{skipper} for an example).
We perform this noise measurement averaging the first $N$ samples of each overscan pixel, where $N$ ranges from $1$ to $400$. 
The resulting noise performance as a function of the number of averaged samples is shown in the left panel of \figref{skipper}.
We measure a single-sample readout noise of $\sigma_{1} = 3.57 \ermspix$, which is consistent with the reported performance of Skipper CCDs elsewhere \citep{Tiffenberg:2017,Rodrigues:2020}. 
We find that the noise performance follows the theoretical expectation from \eqnref{noise}, achieving a readout noise of $\sigma_{400} = 0.18 \ermspix$ for 400 samples (\tabref{performance}). 

\noindent {\it Single-electron resolution and gain:} 
The right panel of \figref{skipper} shows a histogram of pixel values calculated from the average of 200 measurements of each pixel. 
Peaks in this distribution correspond to quantized charge collection demonstrating single-electron/single-photon sensitivity.
From the width of these peaks, we calculate a readout noise of $\sigma_{200} = 0.245 \ermspix$. 
The CCD gain constant, $K$, can be calculated directly from the separation between peaks, since adjacent peaks differ in collected charge by a single electron.
We measure a gain of $K=162 \ADU/\e$ from the average peak-to-peak separation.

\noindent {\it Readout time:} 
The expected readout time can be calculated from clocking frequency combined with the delays and integration times used when reading the CCD. 
We used a sample integration time of $21\us$, since larger integration times are found to yield minimal gains in performance \cite{Cancelo:2020}.
For a large numbers of samples, the readout time is dominated by the time per sample, while the vertical and horizontal shifts make a fractionally larger contribution to the total readout time for a smaller number of samples.
For a 400-sample image, we calculate an expected readout time of $46.22 \us/\pix/\unit{samp}$ for a single amplifier.
Direct measurements of the readout time during data acquisition give a value of $50 \us/\pix/\unit{samp}$ for a single amplifier.
This value is in good agreement with our expectation and measurements of Skipper CCD readout time in the literature \citep{Barak:2020}.
Accounting for the four amplifiers that read in parallel, the readout time for our Skipper CCD is $12.5 \us/\pix/\unit{samp}$.
Readout time is critically important for astronomical applications, and future work will explore techniques to reduce Skipper CCD readout time.


\subsection{Photon Transfer Curves}
\label{sec:ptc}

The photon transfer curve (PTC) describes the CCD response to uniform illumination and is a common diagnostic for CCD performance \cite{Janesick:2001,Astier:2019}. 
Specifically, the PTC can be used to characterize the readout noise, gain, and saturation level of CCDs. 
The PTC is constructed by comparing the average of pixel values ($\mu_S$) to the standard deviation of pixel values ($\sigma_S$) for a range of illumination levels.
 
The PTC illustrates the performance of the CCD in several characteristic regimes (i.e., Figure 2.2 in \cite{Janesick:2001}).
In the readout noise regime, the variance in the signal is dominated by the readout noise and is largely independent of the signal level. 
In the shot noise regime, the variance in pixel values is dominated by the random arrival of photons and is governed by Poisson statistics with $\sigma_S \propto \sqrt{\mu_S}$ (a slope of 1/2 on a log-log plot). 
In the fixed-pattern noise regime, the total noise is dominated by pixel non-uniformity. 
The noise contribution from pixel non-uniformity is proportional to the signal strength, producing a characteristic slope of unity on the log-log plot.
Finally, saturation of the CCD is seen as an abrupt break from the fixed-pattern noise regime.
At this illumination level, charge spreads between pixels lowering the variance \citep{Janesick:2001,Astier:2019}.

\subsubsection{Data Acquisition}
We illuminate the Skipper CCD at a wavelength of 600\nm as selected by the monochromator (\secref{optical}).
To characterize the CCD over a dynamic range spanning $\roughly 5$ orders of magnitude, we use three different optical filters, each with different throughput. 
These filters vary the power incident on the CCD by $\roughly 3$ orders of magnitude.
An integrating sphere is used to provide uniform illumination of the detector.

A series of exposures were taken for each filter.
Exposure times spanned from $1 \second$ to $128 \second$ and were chosen to yield a range of signal values.
Pairs of exposures were taken for each filter/exposure time combination to enable corrections for pixel non-uniformity \citep{Janesick:2001}. 
The first 50 rows of the CCD were read with 200 samples per pixel to study the multi-sample performance of the Skipper CCD.

\begin{figure}[t!]
    \centering
    \includegraphics[width=1.0\textwidth]{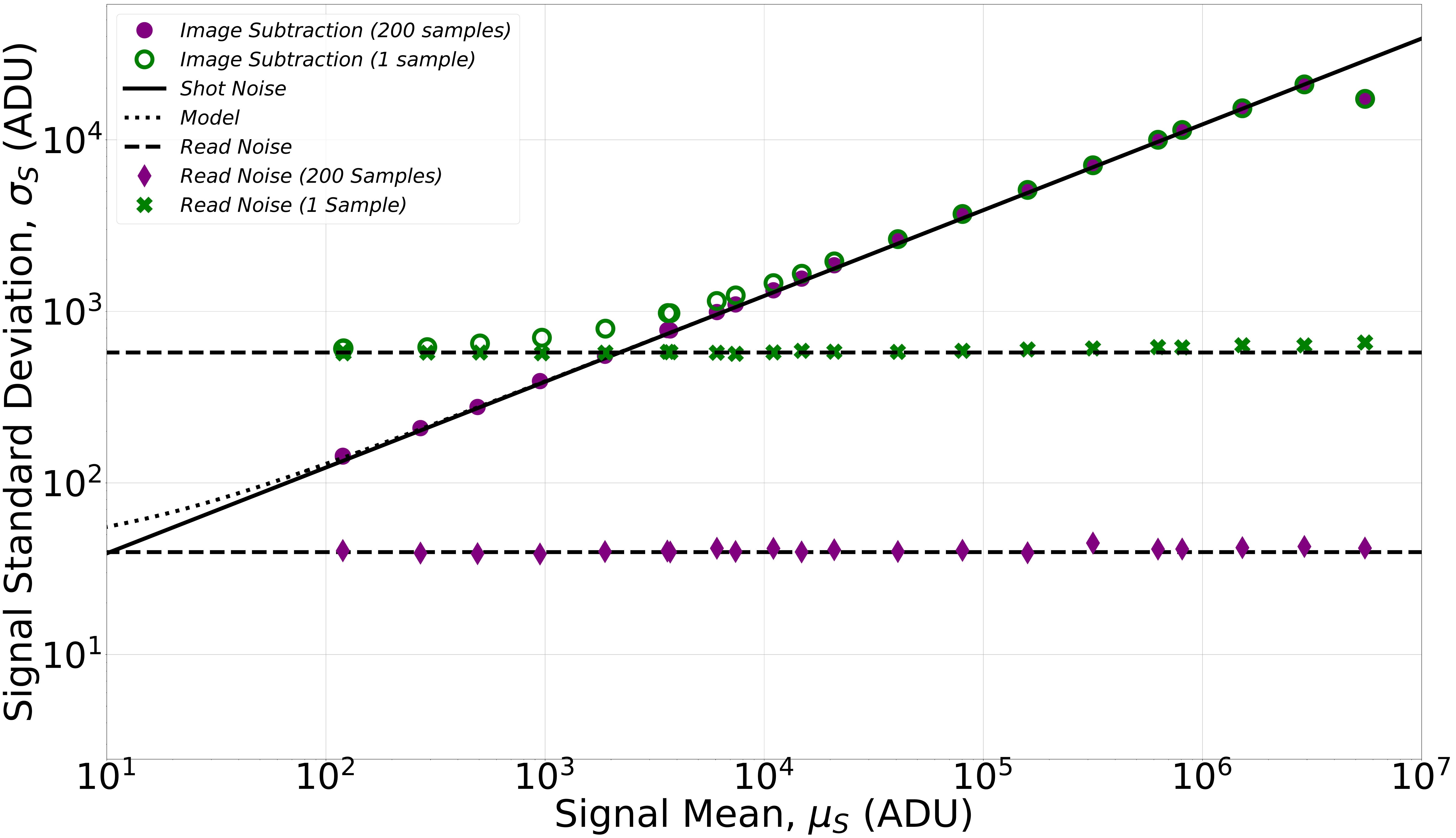}
    \caption{\label{fig:ptc}
    Skipper CCD photon transfer curve (PTC) generated for 1 sample (green) and 200 samples (purple).
    In both data sets, the CCD response is seen to be linear over the shot noise dominated regime.
    The single- and multi-sample data are consistent for signal levels $\gtrsim 2 \times 10^3 \ADU$ ($\gtrsim 12\e$); however, significant deviations are visible at lower signal levels due to the difference in readout noise between the two configurations.
    The multi-sample data show a reduction of readout noise across all signal levels. 
    The apparent increase in the readout noise at signal levels $\gtrsim 3 \times 10^5 \ADU$ ($\gtrsim 1900 \e$) is actually a manifestation of excess charge in the overscan pixels (see text for details).
    }
    
\end{figure}

\subsubsection{Results}

We calculate the mean pixel value from the two exposures taken in each filter/exposure time configuration.
The mean signal level in each image is calculated as
\begin{equation}
    \mu_S = \frac{1}{2}(\mu_{1} + \mu_{2}) = \frac{1}{2 N_{\rm pix}} \sum_i^{N_{\rm pix}} (S_{1,i} + S_{2,i})
\end{equation}
where $i$ indexes the pixels and $N_{\rm pix}$ is the total number of pixels in the image.
$S_{1,i}$ and $S_{2,i}$ are the measured values of pixel $i$ in image 1 and 2, respectively, after mean value of the overscan pixels in the row of pixel $i$ have been subtracted.
In the case of the 200 sample images, the measured values $S_{1,i}$ and $S_{2,i}$ are the average of the individual samples of pixel $i$.
The noise is estimated by subtracting the pair of images, pixel-by-pixel, consequently removing non-uniformity in pixel response caused by fixed-pattern noise. 
The standard deviation of the signal is given by  
\begin{equation}
\sigma_{S} = \frac{1}{2 N_{\rm pix}} \sum_{i}^{N_{\rm pix}} \left(S_{1,i}-\mu_{1}\right)-\left(S_{2,i}-\mu_{2}\right)
\end{equation}
\noindent where $S_{1,i}$, $S_{2,i}$, $\mu_{1}$, and $\mu_{2}$ are the individual pixel values and the mean pixel values in the first and second image, respectively. 
The readout noise is obtained from the overscan pixels in each image following the multi-Gaussian fitting procedure described in \secref{readnoise}.

\noindent {\it PTC measurement:} \figref{ptc} shows the PTC over the dynamic range of the Skipper CCD. 
We show curves derived from both 1-sample and 200-sample flat field images to demonstrate the reduction in readout noise. 
For this detector, we observe saturation at $\roughly 34{,}000 \e$. 
The full-well capacity of other thick, fully-depleted devices is significantly larger ($>150,000 \e$) \citep{Diehl:2008, Flaugher:2015}, and the low saturation level of our device can be attributed to the smaller voltages that we apply to the clock lines to define the pixel boundaries and shift the charge through the horizontal register.
This lower potential was optimized to reduce spurious charge generation for rare particle searches \cite{Tiffenberg:2017} and can be adjusted for applications that require larger full-well capacity.

\noindent {\it Readout noise:} We note a deviation in the readout noise behavior for the 200-sample image at a signal value of  $\mu_{S} \approx 3.2 \times 10^5 \ADU$ ($\approx 1{,}900 \e$).
The change in readout noise corresponds to an increase in charge collection in the overscan pixels, which may result from strong illumination of the serial register during the exposure. 
Rodrigues et al. \cite{Rodrigues:2020} placed a copper shield over the serial register of their Skipper CCD and found no indication of increased charge collection in the overscan. 
We reanalyze their data and confirm that for illumination levels of $\mu_{S} \roughly 400 \e$ there is no indication of charge collection in the overscan; however, further tests should be performed at higher illumination levels. 
For our images, the charge collected in the overscan combined with row-to-row shifts in the bias level has the effect of blurring single-electron peaks when the pixel value distributions for multiple rows are examined.
We confirm that we can retain the same readout noise at high illumination levels by performing row-to-row bias corrections.

\noindent {\it Gain:} The CCD gain  can be derived from the PTC by fitting the slope of the linear relationship between signal variance ($\sigma_{S}^{2}$) and signal mean ($\mu_{S}$) \citep{Janesick:2001}.
From the PTC in \figref{ptc}, we derive a gain of $K=163 \ADU/\e$, which agrees with the gain calculated from the peak-to-peak separation of individual electrons to within 0.6\%. 

\noindent {\it Charge transfer inefficiency:} We measure the charge transfer inefficiency (CTI) in a 200-sample image using the extended pixel edge response (EPER) method \cite{Janesick:2001}. 
Briefly, we are interested in measuring the excess counts in the first overscan column resulting from CTI.
This value is normalized to the total number of electrons in the last physical column and the number of horizontal charge shifts to calculate the CTI.
We calculate this value from several illuminated exposures from the PTC ({$\mu_{S} \sim 37 \e$} to {$\mu_{S} \sim 3820 \e$}) and find results ranging from $3.5 \times 10^{-7}$ to $3.8 \times 10^{-6}$, with the mean value of $1.9 \times 10^{-6}$.
We note that this conventional EPER calculation only accounts for the horizontal charge shifts and not the 200 samples performed during readout. 


\subsection{Relative Quantum Efficiency}
\begin{figure} [!t]
   \begin{center}
   \includegraphics[width=0.85\textwidth]{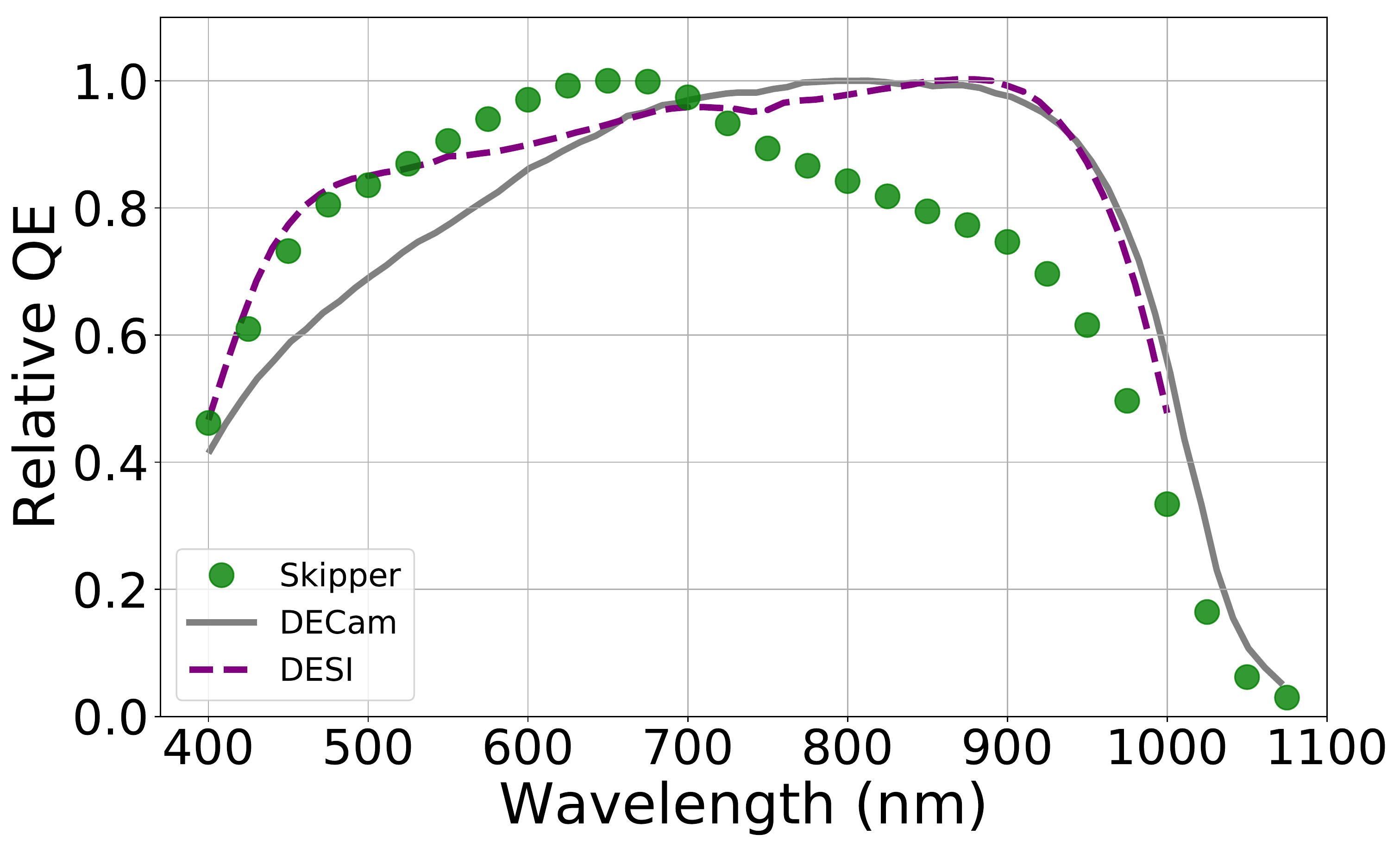}
   \end{center}
   \caption[example] 
   {\label{fig:qe} Relative quantum efficiency of Skipper CCD (green circles) compared to the relative quantum efficiency of a DECam CCD (gray line) \cite{Flaugher:2015} and a DESI CCD tested at Fermilab (purple dashed line).}
\end{figure}

The quantum efficiency (QE) of a CCD is defined as the efficiency with which the detector converts incident photons into electrons \citep{Janesick:2001}.
While it is easy to determine the number of electrons collected by the Skipper CCD, our optical setup does not provide an accurate measurement of the absolute incident photon flux at the detector. 
Thus, we are forced to measure a relative QE based on the relative illumination of the detector.
We define the relative QE as
\begin{equation}
    \label{eqn:qe}
    \QEr = N_\e \times \left( P \frac{\lambda}{hc} \right)^{-1}.
\end{equation}
In the above equation, $N_\e$ is the mean number of electrons measured by the CCD, and is calculated from the mean signal in an image and the gain, $N_{e^{-}}=K^{-1} \mu_{S}$ (\secref{ptc}).
The second term represents the relative photon flux measured by the photodiode mounted on the integrating sphere, where $P$ is the measured power, $\lambda$ is the wavelength setting of the monochromator, $h$ is Planck's constant, and $c$ is the speed of light.
By scanning the monochromator wavelength setting, we can measure the relative QE as a function of incident photon energy.
It is conventional to normalize the relative QE curve with respect to the largest measured value.

\subsubsection{Data Acquisition}

Using the optical setup described in \secref{optical}, we selected a broad-band filter and scanned the monochromator wavelength selection between $400\nm$ and $1100\nm$ in 20\nm steps. 
For each wavelength setting, we exposed the CCD to light for $5\second$ and read out with a single sample per pixel.
During the exposure, we recorded the power received by the photodiode at the integrating sphere.

\subsubsection{Results}

\noindent {\it QE Curve:} 
In \figref{qe}, we show relative QE curves as a function of wavelength for the Skipper CCD and two conventional scientific CCDs measured with similar optical setups at Fermilab.
The first is a high-resistivity, 250\um thick CCD from DECam \cite{Flaugher:2015} and the second is a 250\um thick DESI CCD \cite{Bebek:2017}.
Our Skipper CCD has similar substrate properties and the same anti-reflective coating as the DESI CCD.
Thus, the difference in red sensitivity is slightly unexpected.
However, we note that a vacuum failure during the COVID-19 shutdown may have allowed some moisture to condense on the surface of our CCD leading to decreased QE.
More precise relative and absolute QE measurements of Skipper CCDs are an interesting avenue for future work.

\subsection{Regional Multi-Sampling}

\begin{figure} [!t]
   \begin{center}
   \begin{tabular}{c} 
   \includegraphics[width=0.45\textwidth]{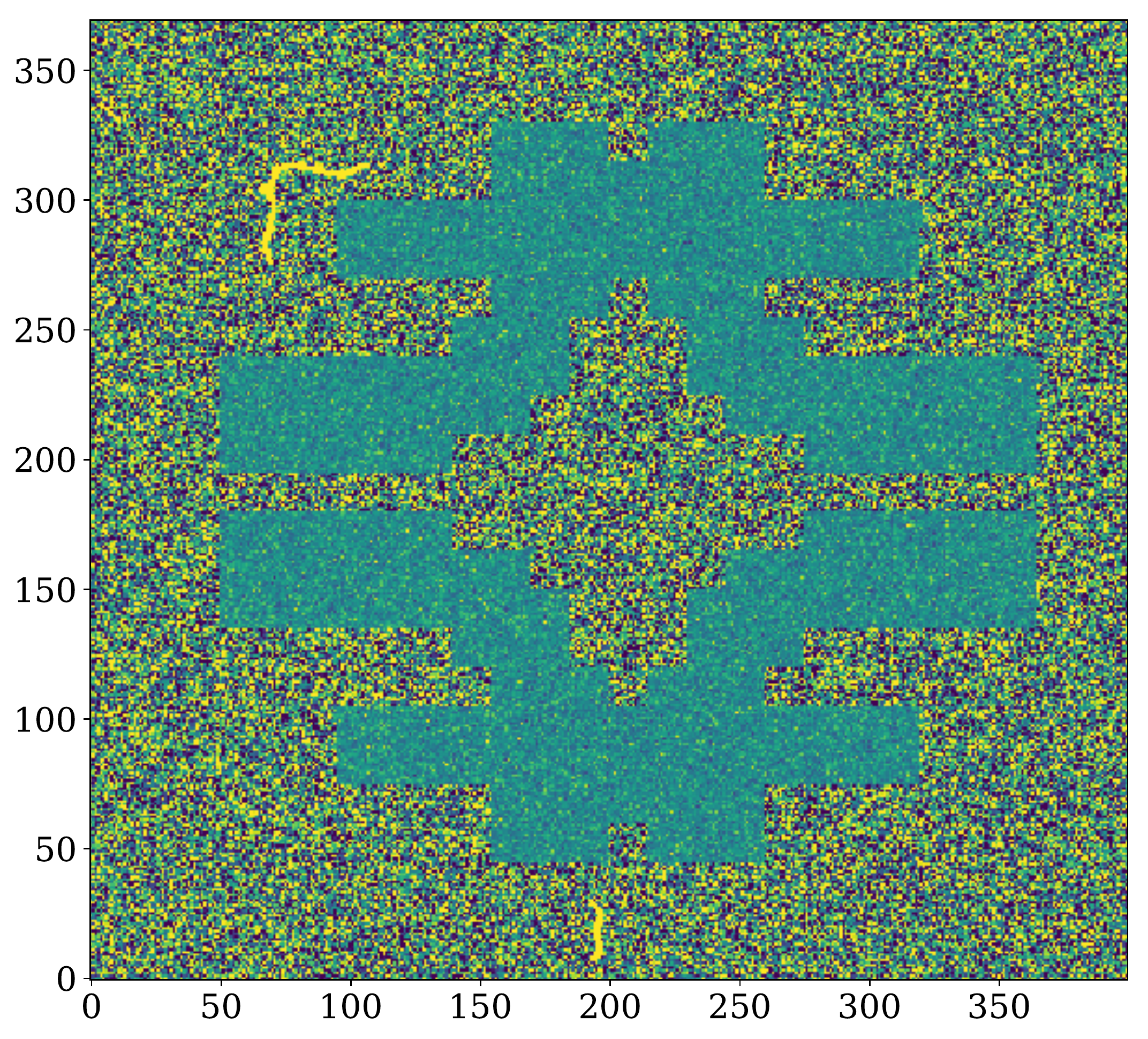}
   \includegraphics[width=0.50\textwidth]{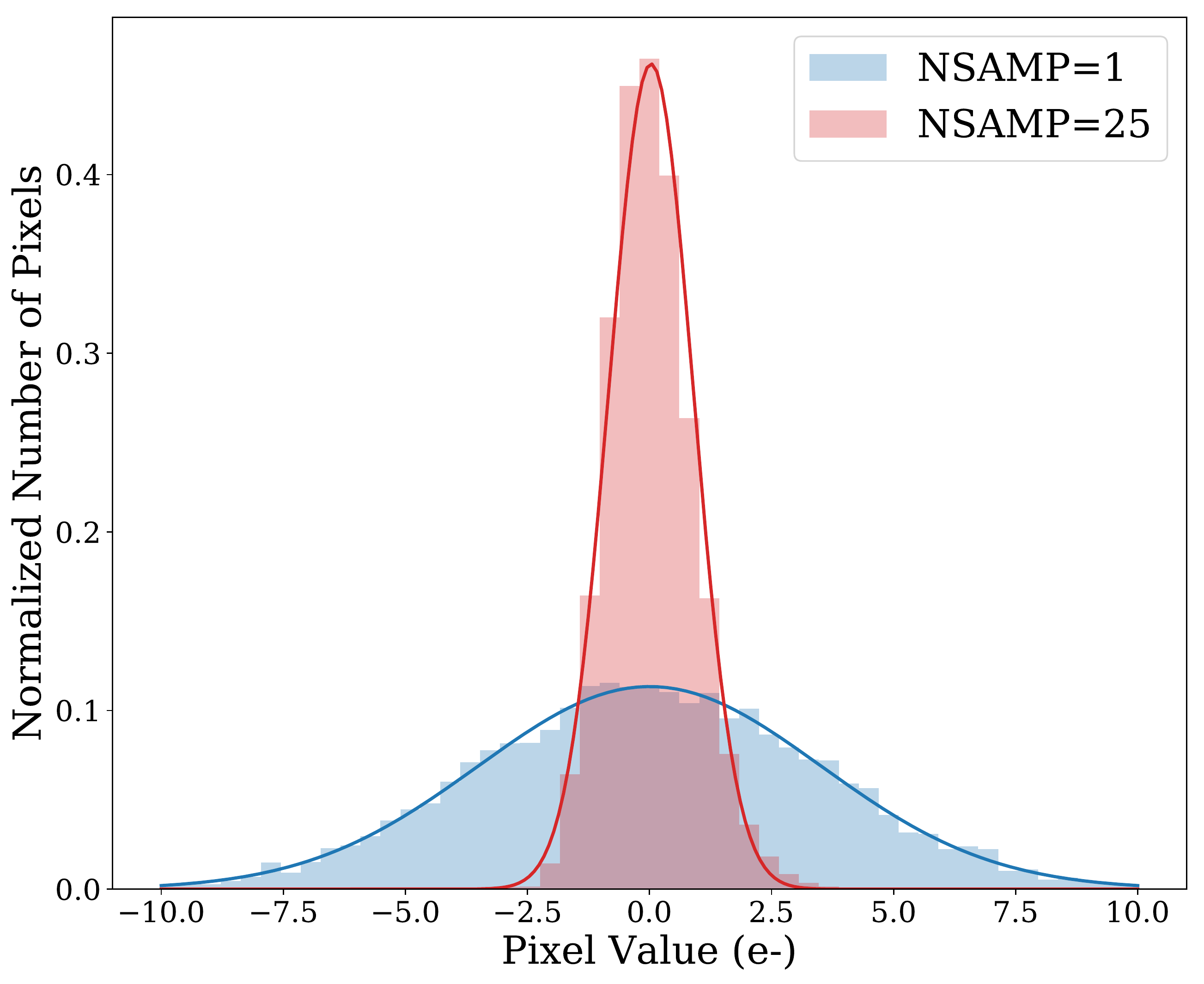}
   \end{tabular}
   \end{center}
   \caption{\label{fig:logo} 
   Example of regional multi-sampling with a Skipper CCD. Left: Spatial distribution of sample-averaged pixel values for a dark exposure. The green-blue region in the shape of the Fermilab logo was read out with 25 samples per pixel, while the outer region was read out with a single sample per pixel. Right: Histogram of pixel counts distribution in the single-sample (blue) and multi-sample (red) regions. The standard deviation of the single sample region is 3.52 \ermspix, while the standard deviation of the multi-sample region is 0.86 \ermspix.
   }
\end{figure}

The Skipper CCD was originally conceived as a technique to read specific pixels of a CCD while skipping other pixels \cite{Janesick:1990,Janesick:2001}.
The new generation of Skipper CCDs make it possible to measure the charge in specific pixels multiple times.
This capability is useful when low-noise performance is necessary only over a specific region of  the CCD. 
This regional selection can significantly reduce readout time when specific regions of interest can be identified prior to readout. 
For example, precursor imaging can provide pixel locations of features of interest (e.g., for direct imaging of exoplanets).
The systemic velocity of resolved stellar systems can provide prior information for elemental line locations when performing spectroscopic observations of faint member stars.
Furthermore, broadband photometric redshifts provide prior information for the location of emission lines when performing spectroscopic observations of distant galaxies.
Motivated by these potential applications, we demonstrate the regional multi-sampling performance of our Skipper CCD.

\subsubsection{Data Acquisition}
To implement regional multi-sampling, we construct a custom readout routine referred to as a ``sequencer'' \cite{Cancelo:2020}. 
We divide the CCD into $15 \times 15 $ pixel blocks, and we specify whether the pixels in each block are measured once or multiple times. 
The geometry of the multi-sampled region and the number of samples per pixel is configurable on an exposure-by-exposure basis. 
We design a sequencer that samples pixels in a complex geometry 25 times, which is sufficient to demonstrate sub-electron noise.
We take 26 exposures in this configuration under very low illumination conditions. 
We calculated the average of the individual measurements for each multi-sampled pixel.

The readout electronics are known to experience a baseline shift when the clocking sequence is changed. 
This baseline shift is stable from image to image, and it can be corrected using a master bias image.
We generate a sigma-clipped master bias frame from the 26 images taken in the multi-sample configuration and use it to correct the baseline shift in each image.

\subsubsection{Results}

\noindent {\it Sub-electron noise:}  
After correcting each frame by the sigma-clipped master bias, we achieve a single-sample readout noise of 3.52 \ermspix and sub-electron readout noise of 0.86 \ermspix in the multi-sample region (\figref{logo}).
Based on the performance demonstrated in \secref{readnoise}, we would expect the multi-sample region to have a readout noise of $\roughly  0.7 \ermspix$.
The difference between the expected and achieved performance likely comes from two factors: 
(1) The presence of low-level illumination broadens the distribution of pixel values by contributing electrons to some pixels. This can be seen to slightly skew the multi-sample distribution to positive values.
(2) Statistical noise in the master bias broadens the pixel-value distribution. Building the master bias from a larger number of input images would reduce this noise.
While these factors are present in both the single- and multi-sample regions, they are most apparent in the multi-sample region due to the reduced readout noise.

The LTA firmware is being actively developed to improve the performance of regional multi-sampling.
Future versions of the firmware will correct for baseline shifts during the readout of individual images.
Furthermore, the LTA will allow the number of samples per pixel to be determined dynamically based on the charge measured in the first read of the pixel.
This will allow Skipper CCDs to operate in a mode where the readout noise is chosen to always be subdominant to the Poisson shot noise in each pixel. 
In principle, a Skipper CCD with very large pixels could have a very large full-well capacity, while also having sensitivity to very small signals by reducing readout noise through multiple measurements. 
Such a detector could provide an extremely large dynamic range.

\section{SUMMARY AND OUTLOOK}

We have presented the first characterization of next-generation Skipper CCDs at optical/near-infrared wavelengths with an emphasis on future applications in astronomy and cosmology.
Starting from a single-sample readout noise of $3.57 \ermspix$, we achieved a readout noise of $0.18 \ermspix$ after 400 non-destructive measurements of the charge in each pixel.
We demonstrated readout noise that scales inversely with the square-root of the number of samples, confirming measurements elsewhere in the literature \cite{Tiffenberg:2017,Rodrigues:2020,Cancelo:2020}.
This detector can achieve sub-electron readout noise with $\roughly 15$ samples per pixel and single photons can be easily resolved with $\roughly 100$ samples per pixel.

We studied the response of Skipper CCDs to optical/near-infrared photons by performing PTC and relative QE measurements. 
In particular, we demonstrated that the detector gain calculated from the PTC agrees with that estimated directly from single-electron peaks to better than 1\%.
We also showed that these detectors achieve a relative QE of $>75\%$ from 450\nm to 900\nm. 
Finally, we demonstrated the capability to reduce readout noise over configurable sub-regions of these detectors.

Long readout time is the largest hurdle faced by Skipper CCDs in astronomy and cosmology. 
The readout time of current Skipper CCDs scales nearly linearly with the number of samples and quadratically with the reduction in readout noise. 
A wide range of hardware, software, and CCD design improvements are being explored to reduce readout time.
These range from conventional solutions (i.e., more readout channels and lower single-sample noise) to more developmental techniques (i.e., firmware developments to dynamically determine the number of samples per pixel based on a predefined charge measurement threshold).
Regional multi-sampling with current detectors already makes it possible to drastically reduced readout noise over a sub-region of a detector while only moderately increasing readout time.

Historically, astronomers have been forced to plan observations such that readout noise is a subdominant contributor to the overall noise budget.
However, as the observational frontier pushes to fainter and more distant objects, the readout noise of conventional CCDs can present a major limitation.
Skipper CCDs offer a mechanism to dramatically reduce readout noise while retaining the well-studied benefits of conventional CCDs.


\acknowledgments 
 
This work was supported in part by Fermilab LDRD 2019.011.  
The work of S.U.\ is supported in part by the Zuckerman STEM Leadership
Program.
CCD development was supported by the Lawrence Berkeley National Laboratory Director, Office of Science, of the U.S.\ Department of Energy under Contract No.\ DE-AC02-05C
H11231. 
This manuscript has been authored by Fermi Research Alliance, LLC under Contract No.\ DE-AC02-07CH11359 with the U.S.\ Department of Energy, Office of Science, Office of High Energy Physics. 
The United States Government retains and the publisher, by accepting the article for publication, acknowledges that the United States Government retains a non-exclusive, paid-up, irrevocable, world-wide license to publish or reproduce the published form of this manuscript, or allow others to do so, for United States Government purposes.

\renewcommand{\refname}{REFERENCES}

\bibliography{main} 
\bibliographystyle{spiebib} 

\end{document}

%% file: commands.tex
\usepackage{amsmath}
\usepackage{amssymb}
\usepackage{xspace}
\usepackage{xifthen}
\usepackage{eso-pic}

\newcommand{\reportnum}[2]{
  \AddToShipoutPictureBG*{%
    \AtPageUpperLeft{%
      \hspace{0.65\paperwidth}%
      \raisebox{#1\baselineskip}{%
        \makebox[0pt][l]{\textnormal{#2}}
  }}}%
}

\mathchardef\mhyphen="2D

\newcommand{\roughly}{\ensuremath{ {\sim}\,} }

\newlength{\dhatheight}

\newcommand{\unit}[1]{\ensuremath{\mathrm{\,#1}}\xspace}

\newcommand{\angstrom}{{\unit{\AA}}}
\newcommand{\nm}{\unit{nm}}
\newcommand{\um}{\unit{\mu m}}
\newcommand{\cm}{\unit{cm}}

\newcommand{\second}{\unit{s}}
\newcommand{\us}{\unit{\mu s}}

\newcommand{\e}{\unit{e^{-}}}

\newcommand{\pix}{\unit{pix}}

\newcommand{\rmspix}{\unit{rms/pix}}
\newcommand{\ADU}{\unit{ADU}}
\newcommand{\ermspix}{\e \rmspix}
\newcommand{\epixday}{\unit{\e \pix^{-1} \unit{day}^{-1}}}
\newcommand{\epixsec}{\unit{\e \pix^{-1} \second^{-1}}}

\newcommand{\texp}{\ensuremath{t_{\rm exp}}\xspace}
\newcommand{\tobs}{\ensuremath{t_{\rm obs}}\xspace}
\newcommand{\tread}{\ensuremath{t_{\rm read}}\xspace}

\newcommand{\QEr}{\ensuremath{\rm QE_{rel}}\xspace}

\newcommand{\Lya}{{Lyman-$\alpha$}\xspace}
\newcommand{\SNR}{\ensuremath{\rm S/N}\xspace}

\newcommand{\secref}[1]{Section~\ref{sec:#1}}

\newcommand{\tabref}[1]{Table~\ref{tab:#1}}
\newcommand{\figref}[1]{Figure~\ref{fig:#1}}
\newcommand{\eqnref}[1]{Equation~\eqref{eqn:#1}}